\documentclass[aps,prb,reprint,superscriptaddress,amsmath,amssymb]{revtex4-2}

\usepackage{graphicx}
\usepackage{dcolumn}
\usepackage{bm}
\usepackage{hyperref}
\usepackage{xcolor}
\usepackage{physics}
\usepackage{microtype}
\usepackage{float}
\usepackage{placeins}

\hypersetup{colorlinks=true,linkcolor=blue!60!black,
            citecolor=green!50!black,urlcolor=blue!60!black}

\graphicspath{{figures/}}

\begin{document}

\title{Interactions Protect Periodically Driven Time Crystals Against
Dephasing but Destabilise Multipolar-Driven Order}

\author{Karthikeya Machiraju}
\email{karthikeyamachiraju005@gmail.com}
\affiliation{Department of Computer Science and Engineering (AI \& ML), PES University, Bengaluru 560085, India}
\author{Kaustav Bhowmick}
\email{kaustavbhowmick@pes.edu}
\affiliation{Department of Electronics and Communication Engineering, PES University, Bengaluru 560085, India}

\begin{abstract}
We show that stronger interactions protect periodically driven
time-crystal order against dephasing but destabilise order under
random multipolar drives, and that coordination matters beyond the
total interaction strength. We studied a disordered kicked Ising
model of $N=12$ and $16$ spins with imperfect $\pi$ pulses and
per-site dephasing by exact diagonalisation, comparing a periodic
drive, random multipolar drives of order $n=0,1,2$ and the
Thue-Morse sequence, and separated coupling strength from
coordination by comparing a grid with a ring at equal total
interaction strength $zJ$. For the periodic drive the coherence time
scales approximately as $1/\gamma$, and stronger coupling reduces the
transverse tilt on which dephasing acts. For multipolar drives of
order $n\ge1$, doubling the coupling on a ring shortens the coherence
time to $0.54$ ($n=1$), $0.37$ ($n=2$) and $0.44$ (Thue-Morse) at
$N=16$, while random signs ($n=0$) are unaffected. A disorder scan
supports a toggling-frame picture in which interactions spoil the
block cancellation of the pulse error. At equal $zJ$ the grid
outlives the ring by $2.15$ ($N=12$) and $2.78$ ($N=16$) for the
periodic drive despite an identical transverse tilt, a coordination
effect beyond this picture. All ratios hold under three definitions
of the coherence time.
\end{abstract}


\maketitle

\section{Introduction}
\label{sec:intro}

A periodically driven many-body system generically absorbs energy and
heats to infinite temperature~\cite{eckardt2017,bukov2015}.
Subharmonic order, a discrete time
crystal~\cite{wilczek2012,else2016,yao2017,elsereview2020,zaletel2023},
survives because disorder can localise the system and stabilise the
order~\cite{khemani2016,vonkeyserlingk2016}, or because heating is
exponentially slow at high drive frequency, so that an approximately
conserved effective Hamiltonian stabilises the order for long
times~\cite{abanin2015,mori2016bound,kuwahara2016,abanin2017,machado2019,machado2020,luitz2020,else2017prethermal}.
Signatures have been observed in superconducting
qubits~\cite{mi2022}, trapped ions~\cite{zhang2017,kyprianidis2021},
nitrogen-vacancy centres~\cite{choi2017} and nuclear
spins~\cite{beatrez2023,stasiuk2023}. Real devices are open. Coupling
to an environment is expected to destroy the order of a disordered
one-dimensional system at long times~\cite{lazarides2017}, and driven
dissipative variants have been studied~\cite{gong2018}. The coherence
time actually observed is therefore set by the interplay of the drive,
the interactions and the environment.

Beyond disorder and fast driving, a second route to long-lived order replaces the periodic drive by a structured aperiodic one. In random multipolar driving (RMD) of order
$n$, the drive sign is built from blocks whose sums cancel over aligned
windows, which suppresses the low-frequency drive noise. For generic
many-body Hamiltonians the lifetime of the slowly heating regime then
grows with the drive frequency with exponent $2n+1$, and the Thue-Morse sequence is the
$n\to\infty$ limit~\cite{zhao2021rmd,mori2021bounds}. Related
quasiperiodic and aperiodic drives have been studied in many-body
systems~\cite{zhao2019spirals,else2020,dumitrescu2022,nandy2017}, a
rondeau time crystal has been observed in a nuclear-spin
system~\cite{zhao2023rondeau,moon2025rondeau}, and slow heating under RMD
has been demonstrated on a 78-qubit processor~\cite{liu2026rmd}. The
cancellation of pulse errors by sequence design is also the working
principle of dynamical
decoupling~\cite{viola1999,khodjasteh2005,hayes2011}, and its
sensitivity to interactions is a known limitation of robust
Hamiltonian engineering in interacting spin
ensembles~\cite{choi2020}.

Dimensionality and coordination matter for the periodic case.
Ref.~\cite{else2017prethermal} argues that in one dimension with
short-range interactions time-crystalline order at nonzero effective
temperature has a finite correlation time and not an exponentially long
one. Our initial state is fully polarised, which for $J>0$ has the maximal
energy of the Ising term and is therefore an extremal state of the
effective Hamiltonian, so we use this statement only as context, since it does not predict the size of any gain from
coordination in a finite, dephased system. Classical analogues show a
comparable interplay between dimensionality and interaction
range~\cite{pizzi2021classical,ye2021classical}.

\begin{figure*}[t]
  \centering
  \includegraphics[width=\linewidth]{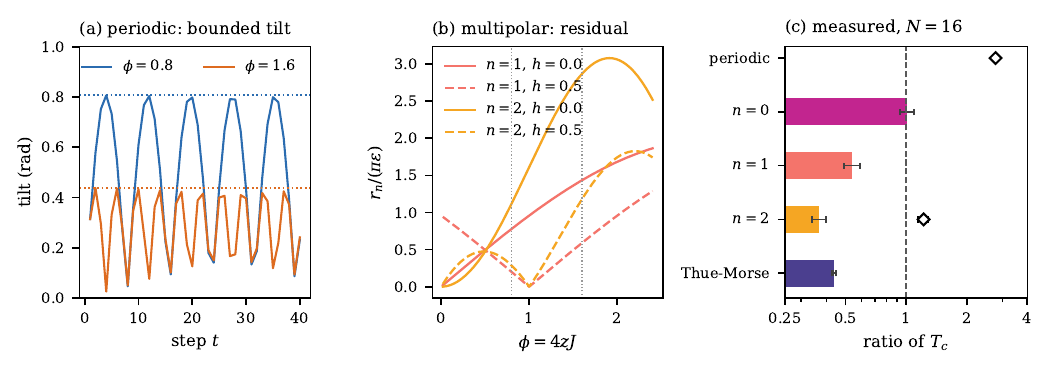}
  \caption{The two regimes. (a) Frozen-neighbour tilt of the periodic drive,
    $|\sum_k\pi\varepsilon\,e^{i(k-1)\phi}|$, for $\phi=0.8$ (ring at $J=0.1$)
    and $\phi=1.6$ (ring at $J=0.2$ and grid at $J=0.1$) with
    $\varepsilon=0.1$; dotted lines are the bound of Eq.~\eqref{eq:floq_bound}.
    (b) First-order block residual $r_n/(\pi\varepsilon)$ of
    Eq.~\eqref{eq:block_resid} for $n=1,2$ and disorder offset $h=0$ and
    $0.5$; dotted verticals mark $\phi=0.8$ and $1.6$. (c) Measured ratios at
    $N=16$ (48 seeds, last-exceedance definition): bars are ring at $J=0.2$
    over ring at $J=0.1$ for $n=0,1,2$ and Thue-Morse; diamonds are grid over
    ring at $J=0.2$, equal $zJ$, for the periodic drive and $n=2$.}
  \label{fig:regimes}
\end{figure*}

This paper asks how the coherence time of each drive responds to the
interaction strength, and whether coordination matters at fixed total
interaction strength $zJ$, with $z$ the coordination number and $J$ the
bond coupling. Chains and square lattices differ in both $z$ and $zJ$,
so the two must be separated explicitly. We did this with exact diagonalisation (ED) of a disordered kicked spin-$1/2$ model with
imperfect $\pi$ pulses and per-site dephasing, on a ring ($z=2$) and a
periodic grid ($z=4$) for $N=12$ and $16$ spins, with a ring at doubled
coupling as the fixed-$zJ$ control. The coherence time $T_c$ is the last
time the autocorrelation magnitude exceeds $1/e$ of its initial value.
Because this is sensitive to late excursions above threshold, we checked every ratio under first-crossing and seed-averaged-trace definitions. The periodic side is the standard rigidity of a discrete
time crystal, recast here as protection against dephasing, and the
multipolar side is consistent in direction with the known algebraic growth
of the lifetime with the drive frequency~\cite{zhao2021rmd}. What is new is the contrast between
the two regimes in one model, the empirical tilt scaling that organises the
periodic drive, and the coordination measurements.  Quantitative claims are those that hold at more than $3\sigma$
under all three definitions at both sizes, and directional claims those
that hold at more than $2\sigma$.
Figure~\ref{fig:regimes} summarises the two regimes and the measured ratios.

The main contributions of this work are the following.
\begin{enumerate}
\item \emph{Opposite response to interactions for different drives}
(Secs.~\ref{sec:dephasing} and~\ref{sec:separation}). Stronger coupling
lengthens the coherence time of the periodic drive under dephasing but
shortens it for multipolar drives of order $n\ge1$. Whether
interactions help therefore depends on the drive design, which matters
for any experiment that uses structured drives.
\item \emph{Separation of coupling strength from coordination}
(Sec.~\ref{sec:separation}). Comparing a grid with a ring at equal $zJ$
isolates the effect of the number of neighbours, which comparisons of
chains and square lattices conflate. The periodic drive shows a
coordination gain that the transverse tilt does not explain, which
identifies an open mechanism.
\item \emph{A mechanism for both regimes} (Secs.~\ref{sec:mechanism}
and~\ref{sec:disorder}). A toggling-frame picture accounts for the protection through a bounded transverse tilt and the destabilisation
through spoiled block cancellation, and a disorder scan matches its prediction that the $n=1$ effect vanishes at strong disorder while the
$n=2$ effect does not.
\item \emph{Robust statistics} (Sec.~\ref{subsec:tc_def}). Every ratio
is reported under three definitions of the coherence time with paired
bootstrap errors, so no conclusion rests on a threshold choice.
\end{enumerate}

Section~\ref{sec:model}
defines the model, methods and the coherence time,
Sec.~\ref{sec:mechanism} gives the picture, Sec.~\ref{sec:dephasing} the dependence on
dephasing, Sec.~\ref{sec:separation} the ring-doubling and fixed-$zJ$
measurements, and Sec.~\ref{sec:disorder} the disorder scan.

\section{Model and Methods}
\label{sec:model}

\subsection{Model}

We studied $N$ spin-$1/2$ sites, initially all up. One step applies the
map of Eq.~\eqref{eq:step}, a rotation about $y$ by $\pi g s_t$ on every
site, then a layer diagonal in the $Z$ basis with the quenched fields
$h_i$ and the Ising bond phase $2J$ on every nearest-neighbour bond. The
pulse amplitude is $g=0.9$, so the pulse error is $\varepsilon=1-g=0.1$;
the fields are drawn uniformly from $[-D/2,D/2]$ with $D=1.5$ and held
fixed per realisation; and every site is dephased after each step by the
channel $\rho\to(1-\gamma)\rho+\gamma Z\rho Z$, with $\gamma=0.01$ unless
stated. The observable is
\begin{equation}
  C(t)=\frac1N\sum_{i=1}^{N}\overline{\langle Z_i(0)\rangle\langle Z_i(t)\rangle},
  \label{eq:autocorr}
\end{equation}
with the overline the average over disorder seeds. Because the $\pi$
pulse reverses every spin, $C(t)$ alternates in sign, and we use $|C(t)|$
at even steps. The next subsection describes the two lattices on which the model was simulated.

\subsection{Lattices}

We compared two lattices at each size, the periodic ring ($z=2$) and a periodic grid ($z=4$), which is the $3\times4$ torus at $N=12$ and the
$4\times4$ torus at $N=16$. The $3\times4$ torus contains $4$ triangles,
one per column, because its three-site direction is periodic. The
$4\times4$ torus contains none, as do the ring and the two-leg ladder. The ladder ($z=3$) was used only in Sec.~\ref{sec:separation}, where it gave
fixed-$zJ$ comparisons at $N=8$, $10$ and $12$. A ring at doubled
coupling is the fixed-$zJ$ control for the grid, since the ring at $J=0.2$ and the grid at $J=0.1$ share $zJ=0.4$. The next subsection defines the drive sequences applied to the pulse error.

\subsection{Drive sequences}

The five sign sequences follow Zhao \textit{et al.}~\cite{zhao2021rmd}. A
$0$-multipole is a random sign, an $m$-multipole is the concatenation of
an $(m{-}1)$-multipole and its negation,
\begin{equation}
  M_m=\bigl(M_{m-1},\,-M_{m-1}\bigr),\qquad |M_m|=2^m,
  \label{eq:rmd_recursion}
\end{equation}
and an order-$n$ RMD is an i.i.d.\ random concatenation of
$n$-multipoles. The $n=2$ multipoles are therefore $(+,-,-,+)$ and
$(-,+,+,-)$. Every aligned block of length $2^k$ with $1\le k\le n$ sums
to zero,
\begin{equation}
  \sum_{t=0}^{2^k-1}s_{\,mT_b+j2^k+t}=0,\qquad 1\le k\le n,
  \label{eq:rmd_constraint}
\end{equation}
where $T_b=2^n$. The five drives are the periodic drive ($s_t=+1$),
$n=0$ (i.i.d.\ random signs), $n=1$, $n=2$ and Thue-Morse
$s_t=(-1)^{s_2(t)}$ with $s_2(t)$ the binary digit sum of $t$, which plays
the role of $n\to\infty$. Section~\ref{subsec:tc_def} defines the coherence time used to compare them.

\subsection{Coherence time}
\label{subsec:tc_def}

The coherence time $T_c$ is the last time $|C(t)|$ exceeds $1/e$ of its
initial value $|C(0)|$, taken over even steps. It equals the time obtained
from a persistence envelope, the first time a trailing $20$-step running
maximum of $|C|$ falls below $1/e$ and stays below, minus the $20$-step
window. On all $1392$ traces of this work the two agree exactly. We used the last exceedance because $|C(t)|$ often returns above $1/e$ after first crossing it. A late revival occurs in $68\%$ of our $N\le12$ traces at
$\gamma=0.01$ and moves $T_c$ by a median of $12$ steps, so a definition
based on a single crossing is noise sensitive. To show that no conclusion
rests on this choice, we also computed every ratio with the first crossing of $1/e$ per seed, and with the seed-averaged trace
$\overline{|C|}$. Section~\ref{sec:separation} gives every ratio under all three definitions, with the paired-bootstrap $1\sigma$ and the significance, together with the per-seed
medians. No ratio lies beyond $1\sigma$ on opposite sides of
unity under different definitions; ratios that do not exceed the stated
significance under all three are not claimed in the main text. Section~\ref{subsec:numerics} describes how the traces were computed.

\subsection{Numerical methods}
\label{subsec:numerics}

We simulated the state vector of $2^N$ amplitudes. The rotations were applied as products of single-qubit gates and the diagonal layer as a phase vector. Dephasing was unravelled exactly as quantum jumps, with $Z_i$ applied to site $i$ with probability $\gamma$ per step~\cite{daley2014}, and averaged over trajectories; at $\gamma=0$ the evolution is unitary and
each seed is one deterministic run. Seed $s$ fixed the disorder $h_i$ and, for the random drives, the sign sequence, and the same seeds were used in every cell, so ratios between cells are paired.

We used two implementations. A CPU implementation ran the trajectories sequentially. The cells with $16$ seeds used $30$ trajectories per seed (the coupling sweep used $40$), and the $48$-seed $N=12$ cells used a fixed $50$. A batched GPU implementation, which evolved all trajectories and seeds of a cell together on a graphics processor, was used for the $N=16$ cells with $48$ seeds and $50$ trajectories; the remaining $N=16$ cells ran on the CPU. We validated the GPU implementation against the CPU one.
The unitary evolution agreed to below $3\times10^{-14}$ in $\langle Z\rangle$
($N=12$ ring, $300$ steps, double precision), and at $\gamma=0.01$ for the
ring at $J=0.1$ with $16$ seeds the coherence times were $20.0\,(0.4)$
against $20.4$ for the periodic drive and $99.2\,(10.7)$ against $100.0$
for $n=2$. The two implementations used different random streams for the jumps, so the agreement at $\gamma>0$ is statistical.

The simulation window started at $400$ steps and was doubled until no seed was pinned, that is until every seed had crossed below $1/e$ before the end of the window, up to $6400$ steps ($51200$ for the unitary periodic
drive at $N=12$). Cells that remained pinned are reported as lower bounds.
Uncertainties on means are standard errors over seeds. For the $48$-seed
data, uncertainties on ratios are $1\sigma$ from a paired bootstrap over
seeds ($2000$ resamples, the same resample in both cells), and we quote the
significance $|r-1|/\sigma$; for the $16$-seed data they are quadrature
propagations of the standard errors. A matrix-product-state calculation
for the random drives on a one-dimensional chain of $N=50$ sites is
described in Sec.~\ref{sec:discussion}. Section~\ref{sec:mechanism} gives a toggling-frame picture that organises the measurements.

\section{A Toggling-Frame Picture of the Two Regimes}
\label{sec:mechanism}

This section gives a picture that organises the measurements. It is not
a derivation, its estimates are heuristic and uncontrolled at the
operating point, and its limits are stated at the end. The periodic side
is the standard rigidity of a discrete time
crystal~\cite{else2016,yao2017,khemani2016,vonkeyserlingk2016}, recast
as protection against dephasing. The multipolar side is consistent in direction
with the known algebraic growth of the RMD lifetime with the drive
frequency~\cite{zhao2021rmd}. What we add is the two-regime contrast, the
tilt scaling of Sec.~\ref{subsec:dephasing_law}, and the coordination
measurements.

\subsection{The step map in the frame of the $\pi$ pulses}

Each Floquet step rotates every spin about $y$ by $\pi g s_t$ and then
applies a layer that is diagonal in the $Z$ basis and contains the
quenched disorder and the Ising coupling; dephasing acts in addition and
is treated in Sec.~\ref{subsec:dephasing_law}. With
$R_y(\theta)=e^{-i\theta Y/2}$,
\begin{equation}
  \begin{aligned}
    U_t &= V\,R(s_t),\qquad R(s)=\prod_i R_y^{(i)}(\pi g s),\\
    V &= \exp\Big[-i\Big(\sum_i h_iZ_i+2J\!\!\sum_{\langle ij\rangle}\!Z_iZ_j\Big)\Big],
  \end{aligned}
  \label{eq:step}
\end{equation}
where $s_t=\pm1$ is the drive sign at step $t$. Write
$\pi g s_t=\pi s_t-\pi\varepsilon s_t$ with pulse error
$\varepsilon=1-g$. Rotations about the same axis commute, so
\begin{equation}
  R_y(\pi g s_t)=R_y(\pi s_t)\,R_y(-\pi\varepsilon s_t).
  \label{eq:split}
\end{equation}
Because $R_y(\pm\pi)=\mp iY$, the rotations $R_y(+\pi)$ and $R_y(-\pi)$
are equal up to a global phase, and the sign $s_t$ drops out of the $\pi$ pulse. It survives only in the error rotation, since, up to a phase,
$R(s_t)=P\,E(s_t)$ with $P=\prod_iY_i$ and
\begin{equation}
  E(s)=\exp\Big(i\,\frac{\pi\varepsilon s}{2}\sum_iY_i\Big).
  \label{eq:error}
\end{equation}
The drive sequence therefore acts only on the pulse error.

The pulse operator $P$ reverses every $Z_i$ and commutes with $E(s)$.
Moving all pulses to the left gives
\begin{equation}
  U(t)=P^{\,t}\,V_tE(s_t)\,V_{t-1}E(s_{t-1})\cdots V_1E(s_1),
  \label{eq:toggling}
\end{equation}
with $V_k=V$ for even $k$ and $V_k=\bar V$ for odd $k$, where
$\bar V=PVP$ is $V$ with the sign of every disorder field reversed. The
factor $P^{\,t}$ supplies the period-doubled sign $(-1)^t$ of the
autocorrelation, so $T_c$ measures how long the frame magnetisation
survives. Because the sign of the quenched field alternates from layer to
layer,
\begin{equation}
  \bar V V=\exp\Big(-4iJ\sum_{\langle ij\rangle}Z_iZ_j\Big),
  \label{eq:echo}
\end{equation}
so over every pair of steps the disorder echoes out and only the Ising phase
survives.

The error rotation at step $k$ is not applied to the bare spins.
Commuting the diagonal layers to the left, $E(s_k)$ is conjugated by the
$k-1$ layers before it, which rotates each $Y_i$ about $z$ by
\begin{equation}
  \phi_i^{(k)}=2c_kh_i+4J(k-1)\sum_{j\in\partial i}Z_j,
  \label{eq:axis}
\end{equation}
where $\partial i$ is the set of neighbours of site $i$ and
$c_k=\sum_{m=1}^{k-1}(-1)^m$, so $c_k=0$ for odd $k$ and $c_k=-1$ for
even $k$. The disorder part is a fixed offset of $-2h_i$ on every second
error rotation, and does not grow. The interaction part grows linearly
with $k$ at a rate set by the neighbour sum. For the fully polarised
initial state the neighbour sum equals $z$ and the rate is
\begin{equation}
  \phi=4zJ
  \label{eq:phi}
\end{equation}
per step, the total interaction strength. Section~\ref{subsec:periodic} applies this frame to the periodic drive.

\subsection{Periodic drive: bounded tilt}
\label{subsec:periodic}

For $s_k=+1$ every error rotation has the same sign. If the axes did not
move, the rotations would add coherently and tilt the spins by an angle
that grows as $\pi\varepsilon t$. With the axes precessing at rate $\phi$
the rotations add as vectors of length $\pi\varepsilon$ at angles
$(k-1)\phi$, and to first order in $\varepsilon$ the tilt of a spin in a
frozen neighbour field is bounded,
\begin{equation}
  \Big|\sum_{k=1}^{t}\pi\varepsilon\,e^{i(k-1)\phi}\Big|
  \le\frac{\pi\varepsilon}{|\sin(2zJ)|},
  \label{eq:floq_bound}
\end{equation}
independent of $t$. The bound decreases with $zJ$ for $2zJ<\pi/2$, which
covers every cell studied here ($2zJ\le0.8$). Equivalently, flipping a
spin against its neighbours costs an Ising phase proportional to $zJ$,
which detunes the error rotation from the flip. This is the rigidity of
a discrete time crystal, and Eq.~\eqref{eq:floq_bound} is its
frozen-neighbour form. We checked it on a single spin in a frozen field, in which the $k$-th
error rotation is by $\pi\varepsilon s_k$ about an axis in the $xy$ plane at
angle $(k-1)\phi$, disorder ignored. The largest tilt of the periodic drive
divided by the bound is $0.851$ to $1.000$ for $\varepsilon=0.1$ and
$zJ=0.1$ to $0.8$, and at least $0.998$ for $\varepsilon\le0.01$; it never
exceeds one.

The neighbour sum in Eq.~\eqref{eq:axis} is an operator, not the number
$z$. Away from the polarised state it fluctuates from configuration to
configuration, and the precession is not a single rate. The picture is
therefore $z$-dependent beyond the value of $zJ$ in ways we do not
derive, and the bound is not quantitative. The root-mean-square transverse tilt of Sec.~\ref{subsec:dephasing_law} is $0.44$ to $0.48$ of it. The bound depends on $zJ$ only, so it gives equal tilts for the grid and the ring at $J=0.2$, as measured. Section~\ref{subsec:dephasing_law} adds dephasing to the periodic-drive picture.

\subsection{Dephasing and the tilt scaling}
\label{subsec:dephasing_law}

Per-site dephasing applies $Z_i$ to a spin with probability $\gamma$ per
step. A jump reverses the transverse components $X_i$, $Y_i$ of the Bloch
vector of that spin and leaves $Z_i$ unchanged, so it acts only on the
transverse component. Let $u(t)$ be the bounded transverse displacement
that the drive builds coherently. A jump at time $t_j$ replaces $u(t)$ by
$u(t)-2u(t_j)$ at later times, so each jump leaves a permanent offset of
size up to $2|u(t_j)|$ in a direction that is random because the jump
time is. Offsets from successive jumps add as a random walk, and the
tilt reaches order one after a time inversely proportional to $\gamma$
times the mean square of $u$. We take that mean square from the unitary
plateau,
\begin{equation}
  T_c\simeq\frac{c}{\gamma\,\theta_\perp^{2}},\qquad
  \theta_\perp^{2}=\langle X\rangle^{2}+\langle Y\rangle^{2},
  \label{eq:tilt_law}
\end{equation}
where $\langle X\rangle$ and $\langle Y\rangle$ are single-site
expectation values averaged in square over sites, disorder seeds and even
steps $t\in[100,6400]$ of the unitary ($\gamma=0$) evolution. The
argument ignores the feedback of the offsets on the neighbours. It fixes
the form of Eq.~\eqref{eq:tilt_law}, not the constant $c$, which we measured.

The single-site Bloch vector is shorter than unity because of
entanglement with the other spins. Its mean-square length
$|\vec r|^{2}=\langle X\rangle^{2}+\langle Y\rangle^{2}+\langle Z\rangle^{2}$
on the plateau is $0.2859$, $0.7162$ and $0.2533$ for the ring at $J=0.1$,
the ring at $J=0.2$ and the $3\times4$ grid at $N=12$, and $0.2835$,
$0.6415$ and $0.3916$ for the ring at $J=0.1$, the ring at $J=0.2$ and the
$4\times4$ grid at $N=16$. The quantity $2(1-C_p)$, with $C_p$ the plateau
of $|C(t)|$, counts this shortening as well as the transverse tilt, and
is not the relevant quantity for a process that acts on the transverse
component only. The transverse tilt is given in Table~\ref{tab:tilt}, and
Table~\ref{tab:tiltdata} lists the plateau quantities and the products
$T_c\gamma\theta_\perp^2$ for every cell.

\begin{table}[!htb]
\caption{Mean-square transverse single-site Bloch component
$\theta_\perp^{2}=\langle X\rangle^{2}+\langle Y\rangle^{2}$ on the
unitary plateau (16 seeds, even steps $t\in[100,6400]$), for the ring at
$J=0.1$, the ring at $J=0.2$ and the grid at $J=0.1$ ($3\times4$ at
$N=12$, $4\times4$ at $N=16$).}
\label{tab:tilt}
\begin{ruledtabular}
\begin{tabular}{lccc}
 & ring $J=0.1$ & ring $J=0.2$ & grid $J=0.1$\\
\hline
$\theta_\perp^{2}$, $N=12$ & $0.1279$ & $0.0446$ & $0.0441$\\
$\theta_\perp^{2}$, $N=16$ & $0.1259$ & $0.0423$ & $0.0428$\\
\end{tabular}
\end{ruledtabular}
\end{table}

\begin{table}[!htb]
\caption{Unitary-plateau quantities for the periodic drive ($\gamma=0$, $16$ seeds, even steps $t\in[100,6400]$): transverse tilt $\theta_\perp^2=\langle X\rangle^2+\langle Y\rangle^2$, single-site Bloch length $|\vec r|^2$ (both from even steps), $2(1-C_p)$ and the plateau $C_p$ of $|C(t)|$ (averaged over all steps $t\in[100,6400]$), and the product $T_c\gamma\theta_\perp^2$ at $\gamma=0.0025,0.005,0.01,0.02$ ($N=12$) or $\gamma=0.01$ ($N=16$). Cells are labelled by ring or grid and $J$.}
\label{tab:tiltdata}
\scriptsize
\renewcommand{\arraystretch}{0.9}
\setlength{\tabcolsep}{3pt}
\begin{ruledtabular}
\begin{tabular}{lccccc}
cell & $\theta_\perp^2$ & $|\vec r|^2$ & $2(1-C_p)$ & $C_p$ & $T_c\gamma\theta_\perp^2$ \\
\hline
\multicolumn{6}{l}{\emph{$N=12$}} \\
ring 0.1 & $0.1279$ & $0.2859$ & $1.2892$ & $0.3554$ & 0.031 0.036 0.026 0.047 \\
ring 0.2 & $0.0446$ & $0.7162$ & $0.3623$ & $0.8188$ & 0.062 0.065 0.065 0.069 \\
grid 0.1 & $0.0441$ & $0.2533$ & $1.2242$ & $0.3879$ & 0.101 0.126 0.139 0.148 \\
\multicolumn{6}{l}{\emph{$N=16$}} \\
ring 0.1 & $0.1259$ & $0.2835$ & $1.2387$ & $0.3807$ & 0.0252 \\
ring 0.2 & $0.0423$ & $0.6415$ & $0.4520$ & $0.7740$ & 0.0462 \\
grid 0.1 & $0.0428$ & $0.3916$ & $0.8976$ & $0.5512$ & 0.1294 \\
\end{tabular}
\end{ruledtabular}
\end{table}

The transverse tilt falls with interaction strength on the ring, and is the same for
the grid and the ring at $J=0.2$, which have the same $zJ$.

Equation~\eqref{eq:tilt_law} is an empirical tilt scaling in our data. For the ring
at $J=0.2$ and $N=12$ the product $T_c\gamma\theta_\perp^2$ is $0.062$,
$0.065$, $0.065$ and $0.069$ for $\gamma=0.0025$, $0.005$, $0.01$ and
$0.02$, constant within about $10\%$, and $0.0462$ at $N=16$ and
$\gamma=0.01$. For the ring at $J=0.1$ it lies between $0.026$ and
$0.047$, where $T_c\approx20$ is the time the initial transient takes to reach the plateau,
and for the $3\times4$ grid between $0.101$ and $0.148$, a drift by a
factor $1.5$. The tilt scaling is therefore approximately $1/\gamma$ for the ring
at $J=0.2$ and the grid, but it is not universal across geometries, since the product is $0.1294$ for the $4\times4$ grid at $N=16$ against $0.0462$ for
the ring at $J=0.2$.

At equal $zJ$ the grid and the ring at $J=0.2$ have the same transverse
tilt, so the tilt scaling predicts equal lifetimes. The grid lives longer, by a
factor $1.63$ to $2.17$ over $\gamma=0.0025$ to $0.02$ at $N=12$ ($16$
seeds; $2.15\,(0.01)$ at $\gamma=0.01$ with $48$ seeds) and $2.78$ at $N=16$
and $\gamma=0.01$. This is the
fixed-$zJ$ gain of the periodic drive, and it is therefore not a tilt effect. It enters the product $T_c\gamma\theta_\perp^{2}$ as the
geometry-dependent factor above, not through the tilt. The gain, the ratio of
$T_c$ for the grid and the ring at $J=0.2$ at $N=12$ ($16$ seeds), grows with $\gamma$ and is $1.63$, $1.96$, $2.16$ and $2.17$ for $\gamma=0.0025$, $0.005$, $0.01$ and $0.02$.

The ring at $J=0.1$ has a unitary plateau $C_p=0.355$ at $N=12$, below
$1/e$, and $0.381$ at $N=16$, barely above it. It is therefore not a
locked time crystal, and its $T_c$ is the time the initial transient takes to reach the plateau. Ratios against this cell appear in tables only. Section~\ref{subsec:multipolar} applies the same frame to the multipolar drives.

\subsection{Multipolar drives: cancellation and the disorder offset}
\label{subsec:multipolar}

For an RMD of order $n\ge1$ the sign sequence, built as in
Ref.~\cite{zhao2021rmd}, obeys the block-sum property of
Eq.~\eqref{eq:rmd_constraint}. Within an aligned block of $2^n$ error
rotations with signs $s\,t_k$, $t_k=(-1)^{{\rm popcount}(k-1)}$, the
first-order residual is the vector sum of the rotations with axis angles
$(k-1)\phi+2c_kh$. Because $c_k$ alternates with the parity of $k$,
the sum factorises exactly (checked against the direct first-order vector
sum to $7.5\times10^{-15}$ for $n=1$ to $4$),
\begin{equation}
  r_n\simeq\pi\varepsilon\,\big|1-e^{i(\phi-2h)}\big|
  \prod_{m=1}^{n-1}\big|1-e^{i2^m\phi}\big|.
  \label{eq:block_resid}
\end{equation}
Two cases follow. For $n\ge2$ the residual vanishes as $\phi\to0$ for
any $h$, like $\phi^{\,n-1}$ times $|1-e^{-2ih}|$. For $n=1$ it does not. At $\phi\to0$ it tends to $\pi\varepsilon|1-e^{-2ih}|$, set by the
disorder. Blocks carry independent random signs, so the residuals random
walk and the tilt reaches order one after $B\sim1/r_n^2$ blocks, giving
$T_c\sim2^n/r_n^2$. This agrees in direction, and in the ordering of
the orders, with the algebraic growth of the RMD lifetime with the drive
frequency, exponent $2n+1$, found by Zhao \textit{et al.}~\cite{zhao2021rmd}. The two are scalings in
different variables, the drive frequency there and the interaction phase
$\phi$ here ($T_c\propto\phi^{-2n}$ for small $\phi$), and we do not
compare the exponents.

Without disorder, Eq.~\eqref{eq:block_resid} reduces to
$\pi\varepsilon\prod_{m=0}^{n-1}|1-e^{i2^m\phi}|$, which follows from the
identity $\sum_{k<2^n}t_ke^{ik\phi}=\prod_{m<n}(1-e^{i2^m\phi})$; the identity
holds to a maximum difference of $1.40\times10^{-14}$. The toy block
residual divided by this product is $1.0000$ for $n=1$ and $2$ at $\phi=0.1$
to $1.6$, and at most $1.0003$ for $n=3$, at $\varepsilon=10^{-4}$. For random-sign blocks without disorder at
$\varepsilon=0.1$ the fitted slope of $\log T$ against $\log\phi$ is $-1.97$
for $n=1$ and $-3.93$ for $n=2$, against $-2n$; this is the scaling with the
interaction phase $\phi$, not with the drive frequency.

A stronger interaction raises $\phi$ and with it the residual. Two
consequences of Eq.~\eqref{eq:block_resid} can be tested. First, after
averaging over the disorder the mean square of the first factor is
$2-2\cos\phi\,(\sin D/D)$, so its dependence on $\phi$ has a relative
amplitude $\sin D/D$, which is $0.665$ at $D=1.5$ and $0.047$ at $D=3$.
The first factor is thus independent of $\phi$ once $2h_i$ spans nearly a
full period, which $D=3$ satisfies ($2h_i\in[-3,3]$). The $n=1$ time then
becomes independent of $J$ and its doubling ratio tends to one, while for
$n\ge2$ the remaining factors keep a dependence on $\phi$ and the ratio
stays below one (to first order the $n=2$ value for $\phi=0.8\to1.6$ is
$|1-e^{i1.6}|^{2}/|1-e^{i3.2}|^{2}=0.52$). Second, random signs ($n=0$) have no cancellation to spoil, and
their error random walk does not depend on the axis, so they should be
unaffected. The disorder scan of Sec.~\ref{sec:disorder} tests the
first, and the ring-doubling data test the second. The formula also
predicts that the absolute $n=2$ time falls with $D$ as fast as the
$n=1$ time, through the same first factor. This is not seen. The $n=1$ time on the ring at $J=0.1$ falls from $94.6$ to $33.2$ between $D=0$
and $D=3$, while the $n=2$ time is $76.8$, $91.1$, $100.0$ and $62.4$ at
$D=0$, $0.75$, $1.5$ and $3$. We record this as a failure of the
first-order estimate for the absolute $n=2$ time. The $n=1$ and $n=2$
coherence times are limited by the pulse error and should depend weakly
on $\gamma$; Sec.~\ref{sec:dephasing} shows that they do.

A single-spin toy with the alternating offset and $h$ uniform in
$[-D/2,D/2]$ reproduces the trend in $n=1$, whose doubling ratio rises
from $0.306$ at $D=1.5$ to $0.905$ at $D=3$, but not the ED
value for $n=2$, since the toy ratio rises from $0.179$ to $0.667$ where ED
stays between $0.27$ and $0.38$ (Table~\ref{tab:toy_d}). The toy time at
$D=3$ is only $24$ to $42$ steps, a few blocks, and the toy is a single
run without an error estimate. We regard the toy as consistent with the mechanism for $n=1$ and not informative for $n=2$ at strong disorder. Section~\ref{subsec:tm} turns to the Thue-Morse sequence.

\begin{table}[!htb]
\caption{Single-spin toy with the alternating disorder offset $2c_kh$, $h$ uniform in $[-D/2,D/2]$, random-sign blocks, $300$ realisations, one run without error estimate: time to fall below $1/e$ at $\phi=0.8$ and $\phi=1.6$ (steps) and their ratio.}
\label{tab:toy_d}
\begin{ruledtabular}
\begin{tabular}{lcccc}
Drive & $D$ & $T(0.8)$ & $T(1.6)$ & ratio \\
\hline
$n=1$ & $0.0$ & 70 & 20 & 0.286 \\
$n=1$ & $0.75$ & 128 & 24 & 0.188 \\
$n=1$ & $1.5$ & 124 & 38 & 0.306 \\
$n=1$ & $3.0$ & 42 & 38 & 0.905 \\
$n=2$ & $0.0$ & 72 & 12 & 0.167 \\
$n=2$ & $0.75$ & 132 & 12 & 0.091 \\
$n=2$ & $1.5$ & 112 & 20 & 0.179 \\
$n=2$ & $3.0$ & 36 & 24 & 0.667 \\
\end{tabular}
\end{ruledtabular}
\end{table}

\subsection{Thue-Morse}
\label{subsec:tm}

The Thue-Morse sequence is deterministic. The random-walk argument does
not apply, since successive blocks are not independent, and we give no
estimate of its coherence time. The partial residuals are those of
Eq.~\eqref{eq:block_resid} for every prefix length $2^n$, which suggests
strong cancellation, but the measured behaviour is not predicted by it.
We observe that doubling the coupling on a ring shortens the Thue-Morse
time to $0.44$ at $\gamma=0.01$ and, without dephasing, to between $0.05$ and
$0.18$ depending on the definition of $T_c$, a deeper effect at weaker dephasing that we do not explain. The last subsection states what the picture does and does not say.

\subsection{What the picture does and does not say}

The picture makes qualitative predictions about the response to the
interaction strength, namely a periodic-drive tilt that falls with $zJ$, and a
multipolar residual that rises with it, with a $\phi$-dependence that
survives strong disorder for $n\ge2$ but not for $n=1$, and not at all
for $n=0$. It is not a
derivation. Per pair of steps it compares an Ising phase of $4J$ per bond
with a transverse rotation of $2\pi\varepsilon$, which are comparable at
the operating point ($g=0.9$, $J=0.1$), so no perturbative expansion is
controlled, and the neighbour field is an operator (Sec.~\ref{subsec:periodic}).
In this picture the coupling enters through the precession rate
$4zJ$ of the polarised state, so at fixed $zJ$ it predicts no dependence
on $z$. The fixed-$zJ$ excess of the periodic drive, and that of the
multipolar drives, is not explained by it. Section~\ref{sec:dephasing} tests these predictions against the dephasing data.

\section{Dependence on Dephasing}
\label{sec:dephasing}

\begin{figure}[htb]
  \centering
  \includegraphics[width=\columnwidth]{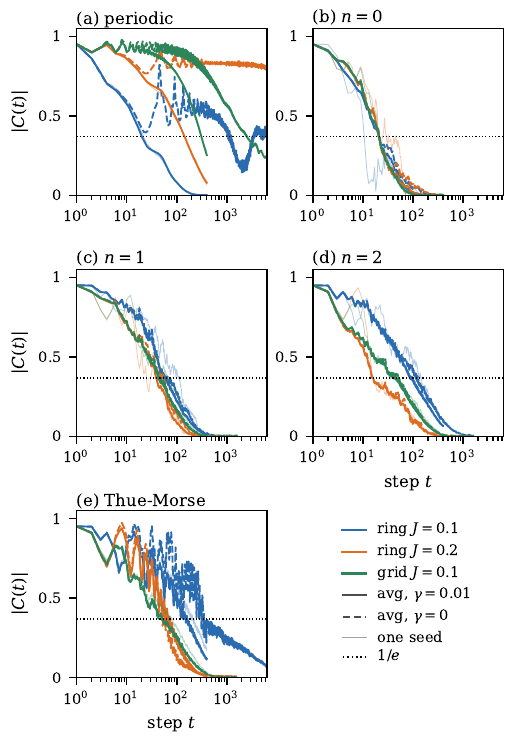}
  \caption{Autocorrelation magnitude $|C(t)|$ at $N=12$ for the five drives
    (a to e): seed averages over $16$ seeds at $\gamma=0.01$ (solid) and
    $\gamma=0$ (dashed) for the ring at $J=0.1$, the ring at $J=0.2$ and the
    $3\times4$ grid at $J=0.1$, and one seed at $\gamma=0.01$ (thin). The
    $\gamma=0.01$ curves end at the simulation window of the cell. Dotted
    line: $1/e$.}
  \label{fig:traces}
\end{figure}

Table~\ref{tab:gamma} gives $T_c$ for the three cells and five drives at
$N=12$ over $\gamma=0$ to $0.02$, and Fig.~\ref{fig:traces} shows the
traces at $\gamma=0.01$ and at $\gamma=0$.

\begin{table}[!htb]
\caption{Coherence time $T_c$ (last exceedance, drive steps) at $N=12$ against the dephasing rate $\gamma$, for the three cells and five drives ($16$ seeds, $30$ trajectories per seed; $\gamma=0$ is unitary). Standard errors in parentheses, none at $\gamma=0$ where each seed is one deterministic run. $^\ddagger$~Lower bound for the periodic drive at $\gamma=0$ (simulation window $51200$ steps, all cells pinned); only the ring at $J=0.2$ has a unitary plateau clearly above $1/e$, so the values for the ring at $J=0.1$ and the grid are threshold crossings of a plateau near $1/e$ and not lifetimes.}
\label{tab:gamma}
\scriptsize
\renewcommand{\arraystretch}{0.9}
\setlength{\tabcolsep}{1.5pt}
\begin{ruledtabular}
\begin{tabular}{lccccc}
 & $\gamma{=}0$ & $0.0025$ & $0.005$ & $0.01$ & $0.02$ \\
\hline
\multicolumn{6}{l}{\emph{Periodic}} \\
ring $J{=}0.1$ & $\geq$50899$^\ddagger$ & 96.8\,(3.7) & 55.8\,(0.7) & 20.4\,(0.5) & 18.4\,(0.3) \\
ring $J{=}0.2$ & $\geq$51179$^\ddagger$ & 559.5\,(9.2) & 289.5\,(5.1) & 145.6\,(2.3) & 77.5\,(1.0) \\
grid $J{=}0.1$ & $\geq$48811$^\ddagger$ & 914.0\,(10.4) & 568.6\,(4.6) & 314.4\,(3.8) & 168.1\,(1.7) \\
\multicolumn{6}{l}{\emph{$n{=}0$}} \\
ring $J{=}0.1$ & 34.0 & 30.8\,(2.9) & 27.2\,(2.7) & 25.8\,(2.7) & 24.1\,(2.4) \\
ring $J{=}0.2$ & 36.4 & 31.0\,(3.8) & 28.6\,(3.1) & 28.4\,(2.9) & 26.0\,(2.1) \\
grid $J{=}0.1$ & 22.8 & 22.9\,(2.3) & 22.5\,(2.3) & 22.5\,(2.3) & 21.2\,(1.8) \\
\multicolumn{6}{l}{\emph{$n{=}1$}} \\
ring $J{=}0.1$ & 94.4 & 81.1\,(9.1) & 74.8\,(7.0) & 70.5\,(6.2) & 63.5\,(5.3) \\
ring $J{=}0.2$ & 57.9 & 52.4\,(7.1) & 47.8\,(6.8) & 44.5\,(5.8) & 39.6\,(4.6) \\
grid $J{=}0.1$ & 52.6 & 51.5\,(5.1) & 51.0\,(5.1) & 47.9\,(4.6) & 43.2\,(4.0) \\
\multicolumn{6}{l}{\emph{$n{=}2$}} \\
ring $J{=}0.1$ & 154.6 & 126.1\,(15.5) & 113.9\,(12.8) & 100.0\,(11.2) & 82.9\,(9.0) \\
ring $J{=}0.2$ & 45.5 & 33.1\,(6.2) & 33.0\,(6.2) & 26.9\,(5.8) & 23.2\,(4.4) \\
grid $J{=}0.1$ & 56.6 & 53.9\,(5.7) & 52.6\,(5.5) & 50.0\,(5.2) & 43.0\,(4.6) \\
\multicolumn{6}{l}{\emph{Thue-Morse}} \\
ring $J{=}0.1$ & 1243.8 & 289.8\,(19.5) & 219.4\,(13.3) & 168.2\,(10.9) & 118.4\,(6.1) \\
ring $J{=}0.2$ & 61.9 & 67.4\,(3.4) & 70.4\,(3.8) & 72.2\,(3.6) & 69.1\,(3.0) \\
grid $J{=}0.1$ & 62.5 & 60.2\,(6.5) & 58.9\,(6.2) & 58.5\,(6.0) & 54.1\,(4.8) \\
\end{tabular}
\end{ruledtabular}
\end{table}

\begin{figure}[htb]
  \centering
  \includegraphics[width=\columnwidth]{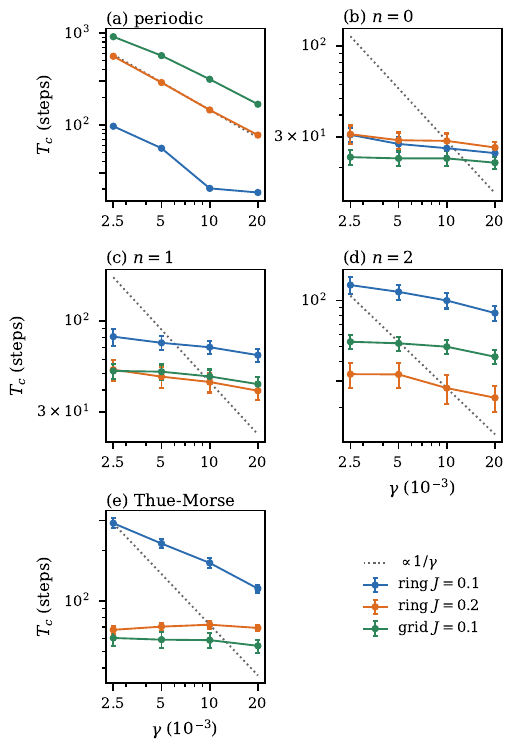}
  \caption{Coherence time $T_c$ against the dephasing rate $\gamma$ at
    $N=12$ for the five drives (a to e), for the ring at $J=0.1$, the ring
    at $J=0.2$ and the $3\times4$ grid at $J=0.1$ ($16$ seeds, $30$
    trajectories; error bars are standard errors). The dotted line is
    proportional to $1/\gamma$ through the ring at $J=0.2$ at $\gamma=0.01$.}
  \label{fig:Tc_gamma}
\end{figure}

\emph{Periodic drive, unitary.} Only the ring at $J=0.2$ has a plateau of
$|C(t)|$ clearly above $1/e$, $0.819$ at $N=12$ and $0.774$ at $N=16$
(Fig.~\ref{fig:traces}a, dashed). Its coherence time is at least
$5.12\times10^{4}$ steps at $N=12$ (mean $51179\,(1)$, $15$ of $16$ seeds
at the window of $51200$) and at least $6380$ at $N=16$, where the window
was $6400$. The ring at $J=0.1$ and the grid have plateaus of $0.355$ and
$0.388$ at $N=12$, close to $1/e$, so their unitary $T_c$ in
Table~\ref{tab:gamma} are threshold crossings of a plateau near $1/e$ and
not lifetimes.

\emph{Periodic drive, dephasing.} With dephasing the periodic-drive time
falls approximately as $1/\gamma$ (Fig.~\ref{fig:Tc_gamma}a). For the ring
at $J=0.2$, $T_c=559.5$, $289.5$, $145.6$ and $77.5$ at
$\gamma=0.0025$, $0.005$, $0.01$ and $0.02$, so that $T_c\gamma$ is
$1.40$, $1.45$, $1.46$ and $1.55$. For the grid $T_c\gamma$ drifts from
$2.29$ to $3.36$ over the same range. The transverse tilt on the unitary
plateau organises this scaling as described in Sec.~\ref{subsec:dephasing_law}. The product $T_c\gamma\theta_\perp^{2}$ is $0.062$ to $0.069$ for the ring at $J=0.2$
and $0.101$ to $0.148$ for the grid (Table~\ref{tab:tiltdata}).

\emph{Multipolar drives.} The $n=1$ and $n=2$ times depend weakly on
$\gamma$ (Fig.~\ref{fig:Tc_gamma}c,d). On the ring at $J=0.1$ the $n=2$
time is $154.6$ at $\gamma=0$ and $82.9$ at $\gamma=0.02$, and the $n=1$
time $94.4$ and $63.5$; in every cell the change over $\gamma=0$ to $0.02$
is less than a factor $2$. The product $T_c\gamma$ is far from constant
(for $n=2$ on the ring at $J=0.1$ it rises from $0.32$ to $1.66$), so
these times are not dephasing limited. The $n=0$ time also changes little
($34.0$ to $24.1$ on the ring at $J=0.1$). Thue-Morse on the ring at
$J=0.1$ is the exception, with $1243.8$ at $\gamma=0$ and $118.4$ at $\gamma=0.02$, a factor $10$. On the other two cells it changes by at
most a factor $1.2$.

\emph{Ratios against $\gamma$.} The ratios of Sec.~\ref{sec:separation}
depend on $\gamma$ as follows (Table~\ref{tab:ratios_gamma}, $16$
seeds). The ring-doubling ratio is nearly independent of $\gamma$ for
$n=1$ ($0.61$ at $\gamma=0$ and $0.62$ to $0.65$ at $\gamma>0$) and $n=2$
($0.29$ at $\gamma=0$ and $0.26$ to $0.29$ at $\gamma>0$), and consistent
with unity for $n=0$ ($1.01$ to $1.10$). For Thue-Morse it depends
strongly on $\gamma$, being $0.05$ at $\gamma=0$ ($0.05$ to $0.18$ across the
three definitions) and $0.23$, $0.32$, $0.43$ and
$0.58$ at $\gamma=0.0025$, $0.005$, $0.01$ and $0.02$; the shortening is
deeper at weaker dephasing. The fixed-$zJ$ gain of the periodic drive in $T_c$
(grid over the ring at $J=0.2$) grows with $\gamma$ ($1.63$, $1.96$, $2.16$ and $2.17$ at $\gamma=0.0025$ to $0.02$) and is not defined at
$\gamma=0$, where all cells are pinned. Section~\ref{sec:separation} uses these ratios to separate the coupling strength from the coordination.

\begin{table}[!htb]
\caption{Ratios of coherence times against the dephasing rate $\gamma$ at $N=12$ ($16$ seeds, $30$ trajectories, unpaired $1\sigma$ in parentheses; $\gamma=0$ is unitary, one trajectory). The periodic-drive entries at $\gamma=0$ are omitted because all cells are pinned at the window.}
\label{tab:ratios_gamma}
\scriptsize
\renewcommand{\arraystretch}{0.9}
\setlength{\tabcolsep}{1.5pt}
\begin{ruledtabular}
\begin{tabular}{lccccc}
 & $\gamma{=}0$ & $0.0025$ & $0.005$ & $0.01$ & $0.02$ \\
\hline
\multicolumn{6}{l}{\emph{ring$_{0.2}$/ring$_{0.1}$}} \\
Periodic & -- & 5.78\,(0.24) & 5.19\,(0.11) & 7.15\,(0.21) & 4.22\,(0.09) \\
$n{=}0$ & 1.07\,(0.20) & 1.01\,(0.16) & 1.05\,(0.15) & 1.10\,(0.16) & 1.08\,(0.14) \\
$n{=}1$ & 0.61\,(0.12) & 0.65\,(0.11) & 0.64\,(0.11) & 0.63\,(0.10) & 0.62\,(0.09) \\
$n{=}2$ & 0.29\,(0.06) & 0.26\,(0.06) & 0.29\,(0.06) & 0.27\,(0.07) & 0.28\,(0.06) \\
Thue-Morse & 0.05\,(0.01) & 0.23\,(0.02) & 0.32\,(0.03) & 0.43\,(0.04) & 0.58\,(0.04) \\
\hline
\multicolumn{6}{l}{\emph{grid/ring$_{0.2}$}} \\
Periodic & -- & 1.63\,(0.03) & 1.96\,(0.04) & 2.16\,(0.04) & 2.17\,(0.04) \\
$n{=}0$ & 0.63\,(0.12) & 0.74\,(0.12) & 0.79\,(0.12) & 0.79\,(0.11) & 0.82\,(0.09) \\
$n{=}1$ & 0.91\,(0.16) & 0.98\,(0.17) & 1.07\,(0.19) & 1.08\,(0.17) & 1.09\,(0.16) \\
$n{=}2$ & 1.24\,(0.24) & 1.63\,(0.35) & 1.59\,(0.34) & 1.86\,(0.44) & 1.85\,(0.40) \\
Thue-Morse & 1.01\,(0.13) & 0.89\,(0.11) & 0.84\,(0.10) & 0.81\,(0.09) & 0.78\,(0.08) \\
\end{tabular}
\end{ruledtabular}
\end{table}

\section{Coupling Strength and Coordination}
\label{sec:separation}

To separate the coupling strength from the coordination number we compared three cells at each size, namely the ring at $J=0.1$, the ring at
$J=0.2$ and the grid at $J=0.1$. The last two share the total interaction
strength $zJ=0.4$. The ratio of the ring at $J=0.2$ to the ring at
$J=0.1$ measures the coupling strength alone. The ratio of the grid to the
ring at $J=0.2$ measures the coordination at fixed $zJ$. The results, from
$48$ seeds at $N=12$ and $N=16$ with the last-exceedance definition, are in
Table~\ref{tab:ratios}; the other two definitions and the significance of every entry are in
Table~\ref{tab:defs}, and the per-seed medians in
Table~\ref{tab:seeds}.

\begin{table}[!htb]
\caption{Ratios of coherence times $T_c$ (last-exceedance definition) between the ring at $J=0.1$, the ring at $J=0.2$ and the grid at $J=0.1$ ($3\times4$ at $N=12$, $4\times4$ at $N=16$), from $48$ disorder seeds with $50$ quantum-jump trajectories each and $\gamma=0.01$. Uncertainties are $1\sigma$ from a paired bootstrap over seeds (the seeds share disorder and drive sequence across cells). The ratio ring$_{0.2}$/ring$_{0.1}$ measures the coupling strength, and grid/ring$_{0.2}$ the coordination at equal $zJ$. $^\dagger$~Periodic-drive ratios against the ring at $J=0.1$ are quoted here only: the unitary plateau of that cell is $0.355$ at $N=12$, below $1/e$, so it is not a locked time crystal and its $T_c$ is the time the initial transient takes to reach the plateau. All three definitions of $T_c$ are in Table~\ref{tab:defs}.}
\label{tab:ratios}
\scriptsize
\renewcommand{\arraystretch}{0.9}
\setlength{\tabcolsep}{2pt}
\begin{ruledtabular}
\begin{tabular}{lccc}
Drive & ring$_{0.2}$/ring$_{0.1}$ & grid/ring$_{0.1}$ & grid/ring$_{0.2}$ \\
\hline
\multicolumn{4}{l}{\emph{$N=12$}} \\
Periodic & 7.31\,(0.13)$^\dagger$ & 15.73\,(0.28)$^\dagger$ & 2.15\,(0.01) \\
$n{=}0$ & 0.99\,(0.07) & 0.85\,(0.05) & 0.85\,(0.06) \\
$n{=}1$ & 0.53\,(0.05) & 0.63\,(0.04) & 1.19\,(0.07) \\
$n{=}2$ & 0.35\,(0.03) & 0.49\,(0.02) & 1.41\,(0.11) \\
Thue-Morse & 0.44\,(0.02) & 0.36\,(0.01) & 0.82\,(0.04) \\
\hline
\multicolumn{4}{l}{\emph{$N=16$}} \\
Periodic & 5.47\,(0.07)$^\dagger$ & 15.22\,(0.19)$^\dagger$ & 2.78\,(0.02) \\
$n{=}0$ & 1.01\,(0.08) & 0.84\,(0.05) & 0.83\,(0.05) \\
$n{=}1$ & 0.54\,(0.05) & 0.63\,(0.03) & 1.16\,(0.07) \\
$n{=}2$ & 0.37\,(0.03) & 0.45\,(0.03) & 1.22\,(0.07) \\
Thue-Morse & 0.44\,(0.01) & 0.35\,(0.01) & 0.80\,(0.03) \\
\end{tabular}
\end{ruledtabular}
\end{table}

\begin{table*}[t]
\caption{Ratios of coherence times under the three definitions of $T_c$ ($48$ seeds, $50$ trajectories, $\gamma=0.01$), $N=12$ (left three columns) and $N=16$ (right three): ratio with paired-bootstrap $1\sigma$ in parentheses, then the significance $|r-1|/\sigma$. First crossing: first even step with $|C|<(1/e)|C(0)|$, averaged over seeds. Last exceedance: last step with $|C|\ge(1/e)|C(0)|$ per seed, averaged over seeds (the definition used throughout). Seed-averaged trace: last exceedance of the seed-averaged $|C|$.}
\label{tab:defs}
\scriptsize
\renewcommand{\arraystretch}{0.9}
\setlength{\tabcolsep}{1.5pt}
\begin{ruledtabular}
\begin{tabular}{lcccccc}
 & \multicolumn{3}{c}{$N=12$} & \multicolumn{3}{c}{$N=16$} \\
Drive & first crossing & last exceedance & seed-avg trace & first crossing & last exceedance & seed-avg trace \\
\hline
\multicolumn{7}{l}{\emph{ring$_{0.2}$/ring$_{0.1}$}} \\
Periodic & 6.93\,(0.13); 47.1$\sigma$ & 7.31\,(0.13); 48.2$\sigma$ & 7.30\,(0.08); 76.0$\sigma$ & 5.23\,(0.06); 66.2$\sigma$ & 5.47\,(0.07); 62.0$\sigma$ & 5.40\,(0.05); 89.8$\sigma$ \\
$n{=}0$ & 0.92\,(0.10); 0.8$\sigma$ & 0.99\,(0.07); 0.1$\sigma$ & 0.91\,(0.10); 0.9$\sigma$ & 0.94\,(0.10); 0.6$\sigma$ & 1.01\,(0.08); 0.2$\sigma$ & 0.91\,(0.10); 1.0$\sigma$ \\
$n{=}1$ & 0.46\,(0.04); 12.5$\sigma$ & 0.53\,(0.05); 9.2$\sigma$ & 0.53\,(0.06); 8.4$\sigma$ & 0.48\,(0.05); 11.4$\sigma$ & 0.54\,(0.05); 9.0$\sigma$ & 0.55\,(0.05); 8.3$\sigma$ \\
$n{=}2$ & 0.22\,(0.02); 33.2$\sigma$ & 0.35\,(0.03); 19.4$\sigma$ & 0.29\,(0.03); 23.6$\sigma$ & 0.23\,(0.02); 34.2$\sigma$ & 0.37\,(0.03); 18.9$\sigma$ & 0.29\,(0.03); 24.2$\sigma$ \\
Thue-Morse & 0.40\,(0.01); 47.7$\sigma$ & 0.44\,(0.02); 37.2$\sigma$ & 0.44\,(0.02); 33.7$\sigma$ & 0.40\,(0.01); 53.9$\sigma$ & 0.44\,(0.01); 38.7$\sigma$ & 0.45\,(0.02); 35.4$\sigma$ \\
\hline
\multicolumn{7}{l}{\emph{grid/ring$_{0.1}$}} \\
Periodic & 14.91\,(0.28); 50.2$\sigma$ & 15.73\,(0.28); 53.3$\sigma$ & 15.60\,(0.14); 104.5$\sigma$ & 14.50\,(0.18); 73.8$\sigma$ & 15.22\,(0.19); 76.7$\sigma$ & 15.10\,(0.09); 155.7$\sigma$ \\
$n{=}0$ & 0.98\,(0.07); 0.3$\sigma$ & 0.85\,(0.05); 3.1$\sigma$ & 0.91\,(0.09); 1.0$\sigma$ & 0.96\,(0.07); 0.5$\sigma$ & 0.84\,(0.05); 3.0$\sigma$ & 0.91\,(0.08); 1.1$\sigma$ \\
$n{=}1$ & 0.71\,(0.04); 6.6$\sigma$ & 0.63\,(0.04); 10.4$\sigma$ & 0.65\,(0.05); 7.0$\sigma$ & 0.70\,(0.05); 6.1$\sigma$ & 0.63\,(0.03); 10.6$\sigma$ & 0.64\,(0.05); 7.5$\sigma$ \\
$n{=}2$ & 0.48\,(0.03); 17.9$\sigma$ & 0.49\,(0.02); 21.5$\sigma$ & 0.48\,(0.03); 18.3$\sigma$ & 0.46\,(0.03); 16.9$\sigma$ & 0.45\,(0.03); 20.1$\sigma$ & 0.42\,(0.03); 16.8$\sigma$ \\
Thue-Morse & 0.38\,(0.01); 56.1$\sigma$ & 0.36\,(0.01); 63.9$\sigma$ & 0.33\,(0.02); 32.0$\sigma$ & 0.38\,(0.01); 77.8$\sigma$ & 0.35\,(0.01); 75.7$\sigma$ & 0.33\,(0.02); 33.8$\sigma$ \\
\hline
\multicolumn{7}{l}{\emph{grid/ring$_{0.2}$}} \\
Periodic & 2.15\,(0.01); 96.3$\sigma$ & 2.15\,(0.01); 100.1$\sigma$ & 2.14\,(0.01); 76.8$\sigma$ & 2.77\,(0.02); 96.5$\sigma$ & 2.78\,(0.02); 108.9$\sigma$ & 2.80\,(0.02); 81.6$\sigma$ \\
$n{=}0$ & 1.06\,(0.07); 0.9$\sigma$ & 0.85\,(0.06); 2.6$\sigma$ & 1.00\,(0.07); 0.0$\sigma$ & 1.02\,(0.05); 0.4$\sigma$ & 0.83\,(0.05); 3.3$\sigma$ & 1.00\,(0.07); 0.0$\sigma$ \\
$n{=}1$ & 1.54\,(0.08); 6.5$\sigma$ & 1.19\,(0.07); 2.6$\sigma$ & 1.22\,(0.10); 2.3$\sigma$ & 1.46\,(0.08); 5.5$\sigma$ & 1.16\,(0.07); 2.4$\sigma$ & 1.17\,(0.08); 2.0$\sigma$ \\
$n{=}2$ & 2.21\,(0.19); 6.4$\sigma$ & 1.41\,(0.11); 3.8$\sigma$ & 1.64\,(0.15); 4.4$\sigma$ & 2.03\,(0.16); 6.4$\sigma$ & 1.22\,(0.07); 3.0$\sigma$ & 1.43\,(0.14); 3.2$\sigma$ \\
Thue-Morse & 0.94\,(0.04); 1.4$\sigma$ & 0.82\,(0.04); 4.6$\sigma$ & 0.75\,(0.06); 4.1$\sigma$ & 0.94\,(0.03); 1.8$\sigma$ & 0.80\,(0.03); 6.3$\sigma$ & 0.73\,(0.06); 4.7$\sigma$ \\
\end{tabular}
\end{ruledtabular}
\end{table*}

\begin{table*}[t]
\caption{Per-seed coherence time $T_c$ at $N=12$ (top) and $N=16$ (bottom) ($48$ seeds): median [interquartile range]; mean (standard error).}
\label{tab:seeds}
\scriptsize
\renewcommand{\arraystretch}{0.9}
\setlength{\tabcolsep}{1.5pt}
\begin{ruledtabular}
\begin{tabular}{lccc}
Drive & ring $J{=}0.1$ & ring $J{=}0.2$ & grid $J{=}0.1$ \\
\hline
\multicolumn{4}{l}{\emph{$N=12$}} \\
Periodic & 20.0 [18.0, 22.0]; 20.0\,(0.4) & 147.0 [140.0, 150.0]; 145.9\,(1.1) & 316.0 [305.5, 322.0]; 314.0\,(1.6) \\
$n{=}0$ & 27.0 [19.5, 32.0]; 26.5\,(1.5) & 23.0 [17.5, 33.0]; 26.2\,(1.8) & 22.0 [17.5, 26.5]; 22.4\,(1.2) \\
$n{=}1$ & 70.0 [51.0, 94.0]; 72.6\,(4.0) & 35.0 [23.5, 50.5]; 38.3\,(3.1) & 44.0 [32.0, 52.5]; 45.5\,(2.5) \\
$n{=}2$ & 97.0 [72.0, 128.0]; 99.0\,(6.1) & 35.0 [19.0, 50.0]; 34.6\,(2.8) & 47.0 [39.5, 60.5]; 48.7\,(2.4) \\
Thue-Morse & 163.0 [130.0, 196.0]; 167.0\,(6.0) & 72.0 [64.0, 82.0]; 72.8\,(2.3) & 50.0 [40.0, 74.0]; 59.8\,(3.4) \\
\hline
\multicolumn{4}{l}{\emph{$N=16$}} \\
Periodic & 20.0 [18.0, 22.0]; 19.9\,(0.3) & 108.0 [105.5, 112.5]; 108.8\,(0.9) & 302.0 [298.0, 308.5]; 302.5\,(1.3) \\
$n{=}0$ & 26.0 [17.5, 32.0]; 25.9\,(1.5) & 22.0 [17.5, 30.5]; 26.2\,(1.8) & 20.0 [16.0, 24.0]; 21.8\,(1.1) \\
$n{=}1$ & 62.0 [48.0, 90.5]; 70.6\,(3.9) & 35.0 [23.5, 50.5]; 38.5\,(3.0) & 42.0 [33.5, 56.0]; 44.6\,(2.5) \\
$n{=}2$ & 100.0 [76.0, 116.5]; 94.6\,(5.1) & 36.0 [22.0, 52.0]; 35.1\,(2.8) & 43.0 [30.0, 52.5]; 42.8\,(2.5) \\
Thue-Morse & 162.0 [132.0, 192.5]; 167.2\,(5.3) & 74.0 [64.0, 80.5]; 74.0\,(2.1) & 50.0 [41.5, 72.0]; 59.0\,(2.9) \\
\end{tabular}
\end{ruledtabular}
\end{table*}

\emph{Coupling strength, multipolar drives.} Doubling $J$ on the ring
shortens the $n=1$, $n=2$ and Thue-Morse times to $0.53\,(0.05)$,
$0.35\,(0.03)$ and $0.44\,(0.02)$ at $N=12$, and $0.54\,(0.05)$,
$0.37\,(0.03)$ and $0.44\,(0.01)$ at $N=16$. Across the three definitions
the $n=1$ ratio is $0.46$ to $0.53$ ($N=12$) and $0.48$ to $0.55$
($N=16$), the $n=2$ ratio $0.22$ to $0.35$ and $0.23$ to $0.37$, and the
Thue-Morse ratio $0.40$ to $0.44$ and $0.40$ to $0.45$; every one is at
least $8.3\sigma$ below unity. Random signs are unaffected, with $0.99\,(0.07)$ and $1.01\,(0.08)$, and at most $1.0\sigma$ from unity under every
definition. The $n=1$ and $n=2$ values are stable in system size
($0.60$, $0.61$, $0.63$ for $n=1$ and $0.25$, $0.26$, $0.27$ for $n=2$ at $N=8$, $10$, $12$ with $16$ seeds) and for pulse amplitudes $g=0.85$ to
$0.95$ (Tables~\ref{tab:sizes} and~\ref{tab:eps}), where the $n=2$ ratio is
$0.29$, $0.27$ and $0.28$; the $n=1$ ratio is $0.64$, $0.63$ and $0.49$
and the Thue-Morse ratio $0.60$, $0.43$ and $0.54$, so for those two the
sign and the order of magnitude are stable but the size is not.

\begin{table}[!htb]
\caption{Dependence on the pulse amplitude $g$ ($\varepsilon=1-g$) at $N=12$, $\gamma=0.01$ ($16$ seeds, $30$ trajectories, unpaired $1\sigma$).}
\label{tab:eps}
\scriptsize
\renewcommand{\arraystretch}{0.9}
\setlength{\tabcolsep}{3pt}
\begin{ruledtabular}
\begin{tabular}{lccc}
Drive & $g=0.85$ & $g=0.9$ & $g=0.95$ \\
\hline
\multicolumn{4}{l}{\emph{ring$_{0.2}$/ring$_{0.1}$}} \\
Periodic & 11.62\,(0.34) & 7.15\,(0.21) & 3.07\,(0.05) \\
$n{=}0$ & 0.95\,(0.15) & 1.10\,(0.16) & 1.25\,(0.19) \\
$n{=}1$ & 0.64\,(0.09) & 0.63\,(0.10) & 0.49\,(0.08) \\
$n{=}2$ & 0.29\,(0.06) & 0.27\,(0.07) & 0.28\,(0.05) \\
Thue-Morse & 0.60\,(0.06) & 0.43\,(0.04) & 0.54\,(0.03) \\
\hline
\multicolumn{4}{l}{\emph{grid/ring$_{0.2}$}} \\
Periodic & 1.88\,(0.07) & 2.16\,(0.04) & 2.45\,(0.03) \\
$n{=}2$ & 1.79\,(0.37) & 1.86\,(0.44) & 1.36\,(0.24) \\
\end{tabular}
\end{ruledtabular}
\end{table}

\begin{table}[!htb]
\caption{Size dependence at $\gamma=0.01$ ($16$ seeds, $30$ trajectories, unpaired $1\sigma$). Top: doubling the coupling on the ring. Bottom: two-leg ladder ($z=3$) at $J=0.1$ against the ring at $J=0.15$, which have the same $zJ=0.3$; the ladder is used only here.}
\label{tab:sizes}
\scriptsize
\renewcommand{\arraystretch}{0.9}
\setlength{\tabcolsep}{3pt}
\begin{ruledtabular}
\begin{tabular}{lccccc}
 & Periodic & $n{=}0$ & $n{=}1$ & $n{=}2$ & Thue-Morse \\
\hline
\multicolumn{6}{l}{\emph{ring$_{0.2}$/ring$_{0.1}$}} \\
$N{=}8$ & 4.97\,(0.16) & 1.11\,(0.16) & 0.60\,(0.10) & 0.25\,(0.06) & 0.42\,(0.05) \\
$N{=}10$ & 4.98\,(0.45) & 1.09\,(0.15) & 0.61\,(0.09) & 0.26\,(0.06) & 0.42\,(0.04) \\
$N{=}12$ & 7.15\,(0.21) & 1.10\,(0.16) & 0.63\,(0.10) & 0.27\,(0.07) & 0.43\,(0.04) \\
\hline
\multicolumn{6}{l}{\emph{ladder$_{0.1}$/ring$_{0.15}$ (equal $zJ$)}} \\
$N{=}8$ & 1.31\,(0.04) & 0.95\,(0.12) & 1.22\,(0.17) & 0.95\,(0.17) & 0.77\,(0.11) \\
$N{=}10$ & 1.55\,(0.03) & 0.97\,(0.12) & 1.20\,(0.13) & 1.01\,(0.16) & 0.83\,(0.09) \\
$N{=}12$ & 1.77\,(0.04) & 1.05\,(0.11) & 1.23\,(0.12) & 1.08\,(0.14) & 0.85\,(0.07) \\
\end{tabular}
\end{ruledtabular}
\end{table}

\emph{Coupling strength, periodic drive.} The ring at $J=0.2$ against the
ring at $J=0.1$ gives $7.31$ at $N=12$ and $5.47$ at $N=16$. As the ring at
$J=0.1$ is not a locked time crystal, this ratio is quoted in
Table~\ref{tab:ratios} only, and the periodic-drive statements of this
paper rest instead on Sec.~\ref{sec:dephasing} and on the coordination
gain below.

\emph{Coordination at fixed $zJ$, periodic drive.} The grid outlasts the
ring at $J=0.2$ by $2.15\,(0.01)$ at $N=12$ and $2.78\,(0.02)$ at $N=16$,
and the value depends on the definition by at most $0.03$ ($2.14$ to
$2.15$ and $2.77$ to $2.80$), more than $75\sigma$ from unity in every
case. The two cells have the same transverse tilt
(Table~\ref{tab:tilt}), so this is not a tilt effect
(Sec.~\ref{subsec:dephasing_law}). It grows with system size from $N=12$
to $N=16$ and with $\gamma$ (Sec.~\ref{sec:dephasing}).

\emph{Coordination at fixed $zJ$, multipolar drives.} For $n=2$ the grid
outlasts the ring at $J=0.2$ at both sizes, with $1.41\,(0.11)$ at $N=12$ and
$1.22\,(0.07)$ at $N=16$ under the headline definition, and $1.41$ to
$2.21$ and $1.22$ to $2.03$ across definitions, at least $3.8\sigma$ and
$3.0\sigma$ from unity. We claim the direction only, since the size
depends on the definition by a factor of about $1.6$. For $n=1$ the direction is
the same, $1.19\,(0.07)$ and $1.16\,(0.07)$ with ranges $1.19$ to $1.54$
($2.3\sigma$ to $6.5\sigma$) and $1.16$ to $1.46$ ($2.0\sigma$ to
$5.5\sigma$); the weakest definition at $N=16$ is at the boundary of our
directional threshold, so we do not claim it. For Thue-Morse the ratio is
below unity under the last-exceedance and seed-averaged definitions
($0.82\,(0.04)$, $0.80\,(0.03)$; $4.1\sigma$ to $6.3\sigma$) but only
$1.4\sigma$ ($N=12$) and $1.8\sigma$ ($N=16$) below under first
crossing, so we make no claim. For $n=0$ the ratio is $0.85\,(0.06)$ and
$0.83\,(0.05)$ under the headline definition but $1.00$ to $1.06$ under the
others, so we make no claim either. At $z=3$ the ladder
gives no such excess for $n=2$ at $N=8$ to $12$ ($0.95$, $1.01$,
$1.08$), and a periodic-drive gain that grows with $N$ ($1.31$, $1.55$,
$1.77$; Table~\ref{tab:sizes}).

\begin{figure}[htb]
  \centering
  \includegraphics[width=\linewidth]{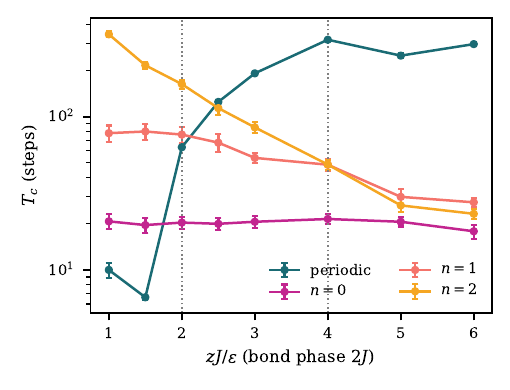}
  \caption{Coherence time against the total interaction strength
    $zJ/\varepsilon$ for four drives ($N=12$, $3\times4$ grid, $\gamma=0.01$,
    $16$ seeds, $40$ trajectories; error bars are standard errors). Dotted
    lines: $zJ/\varepsilon=2$ and $4$, the values of the ring and the grid at
    $J=0.1$ (the sweep itself is on the $3\times4$ grid).}
  \label{fig:sweep}
\end{figure}

\emph{Sweep of the coupling.} Figure~\ref{fig:sweep} sweeps $zJ/\varepsilon$
on the $3\times4$ grid. The periodic-drive time is $10.0\,(1.1)$ and
$6.6\,(0.2)$ at $zJ/\varepsilon=1$ and $1.5$, rises to $63.2\,(1.4)$ at $2$
and $318.2\,(2.3)$ at $4$, and then stays near a plateau of a few hundred
steps ($250.6\,(9.4)$ at $5$ and $298.0\,(9.5)$ at $6$). The point at
$zJ/\varepsilon=4$ was an independent run of the grid cell of Table~\ref{tab:gamma} ($314.4\,(3.8)$) and agreed with it within error. The
$n=2$ time falls from $345.2\,(15.5)$ at $1$ to $23.2\,(1.8)$ at $6$, and the
$n=1$ time is about $78$ up to $zJ/\varepsilon=2$ and falls to
$27.6\,(2.0)$ at $6$. The $n=0$ time stays between $17.9$ and $21.5$. The
drives therefore exchange places. The periodic drive is the shortest-lived
of the four at $zJ/\varepsilon=1.5$, and by $zJ/\varepsilon=2.5$ it exceeds
both $n=1$ and $n=2$ ($125.0$ against $67.8$ and $113.9$). The
weak-coupling dip of the periodic drive is not predicted by the picture of
Sec.~\ref{sec:mechanism}. Section~\ref{sec:disorder} tests the disorder dependence that the picture predicts for the doubling ratio.

\section{Disorder Scan}
\label{sec:disorder}

\begin{figure}[htb]
  \centering
  \includegraphics[width=\columnwidth]{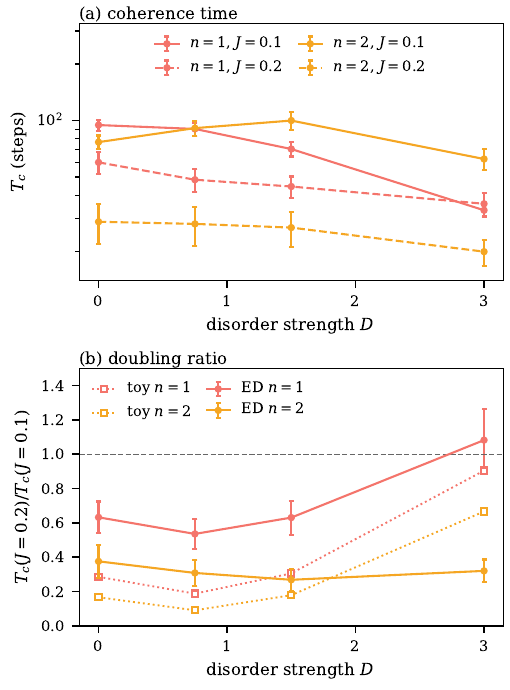}
  \caption{Disorder scan on the ring at $N=12$, $\gamma=0.01$ ($16$ seeds).
    (a) Coherence time for $n=1$ and $n=2$ at $J=0.1$ and $J=0.2$. (b) The
    ratio $T_c(J{=}0.2)/T_c(J{=}0.1)$ from ED (filled, standard errors
    propagated) and from a single-spin toy with the alternating disorder
    offset (open, a single run without error estimate).}
  \label{fig:dscan}
\end{figure}

Equation~\eqref{eq:block_resid} predicts that the dependence of the $n=1$
residual on the interaction phase is lost when the disorder averages the
first factor, and that of $n\ge2$ is not. We tested this on the ring at
$J=0.1$ and $J=0.2$ at $N=12$ with $D=0$, $0.75$, $1.5$ and $3$
($16$ seeds, $\gamma=0.01$). The results are in
Fig.~\ref{fig:dscan} and Table~\ref{tab:dscan}.

\begin{table}[!htb]
\caption{Disorder scan at $N=12$, $\gamma=0.01$ ($16$ seeds, $30$ trajectories): $T_c$ on the ring at $J=0.1$ and $J=0.2$ and their ratio, with standard errors and unpaired $1\sigma$.}
\label{tab:dscan}
\scriptsize
\renewcommand{\arraystretch}{0.9}
\setlength{\tabcolsep}{3pt}
\begin{ruledtabular}
\begin{tabular}{lcccc}
Drive & $D$ & ring $J{=}0.1$ & ring $J{=}0.2$ & ratio \\
\hline
$n{=}1$ & $0.0$ & 94.6\,(6.2) & 59.9\,(7.8) & 0.63\,(0.09) \\
$n{=}1$ & $0.75$ & 90.2\,(7.4) & 48.4\,(6.7) & 0.54\,(0.09) \\
$n{=}1$ & $1.5$ & 70.5\,(6.2) & 44.5\,(5.8) & 0.63\,(0.10) \\
$n{=}1$ & $3.0$ & 33.2\,(2.7) & 36.0\,(5.2) & 1.08\,(0.18) \\
$n{=}2$ & $0.0$ & 76.8\,(6.5) & 28.9\,(6.9) & 0.38\,(0.10) \\
$n{=}2$ & $0.75$ & 91.1\,(8.4) & 28.1\,(6.6) & 0.31\,(0.08) \\
$n{=}2$ & $1.5$ & 100.0\,(11.2) & 26.9\,(5.8) & 0.27\,(0.07) \\
$n{=}2$ & $3.0$ & 62.4\,(8.1) & 20.0\,(3.2) & 0.32\,(0.07) \\
\end{tabular}
\end{ruledtabular}
\end{table}

The $n=1$ doubling ratio is $0.63\,(0.09)$, $0.54\,(0.09)$, $0.63\,(0.10)$
and $1.08\,(0.18)$ at $D=0$, $0.75$, $1.5$ and $3$. The destabilisation by the interaction vanishes at strong disorder, where the first factor of
Eq.~\eqref{eq:block_resid} no longer depends on $\phi$. The $n=2$ ratio is
$0.38\,(0.10)$, $0.31\,(0.08)$, $0.27\,(0.07)$ and $0.32\,(0.07)$, which
stays between $0.27$ and $0.38$ and below one at all $D$, as the surviving
factors predict. The single-spin toy reproduces the $n=1$ rise, from
$0.306$ at $D=1.5$ to $0.905$ at $D=3$, but not the $n=2$ value at strong
disorder, where it rises to $0.667$. The toy times at $D=3$ are $24$ to
$42$ steps, only a few blocks.

The absolute times behave differently from the estimate. The $n=1$ time on
the ring at $J=0.1$ falls from $94.6\,(6.2)$ at $D=0$ to $33.2\,(2.7)$ at
$D=3$, as predicted, but the $n=2$ time is $76.8\,(6.5)$, $91.1\,(8.4)$,
$100.0\,(11.2)$ and $62.4\,(8.1)$, where the same first factor would make it
fall as fast as $n=1$. This is a failure of the first-order estimate for the
absolute $n=2$ time (Sec.~\ref{subsec:multipolar}). The ratio, not the
absolute time, is what the picture predicts. Section~\ref{sec:discussion} discusses what these results mean and where they stop.

\section{Discussion}
\label{sec:discussion}

The sign of the response to interaction strength depends on the drive,
and the two regimes have different origins. For the periodic drive the
coherence time under dephasing is set by the transverse tilt that the pulse
error builds up, and interactions shorten that tilt, in line with the
rigidity of a discrete time crystal~\cite{else2016,yao2017,khemani2016}. This
protection is the effect of dephasing on a tilt, which is why the
unitary lifetime and the dephased lifetime differ by orders of magnitude
(Table~\ref{tab:gamma}). For multipolar drives of order $n\ge1$ the
coherence time is limited by the pulse error and depends weakly on
$\gamma$, and interactions destabilise it because they spoil the
cancellation within blocks, as in the loss of error suppression by
dynamical decoupling sequences in interacting
ensembles~\cite{viola1999,khodjasteh2005,choi2020}. The disorder scan
supports the specific role of the offset $2c_kh$ in this mechanism for
$n=1$, and does not test it for the absolute $n=2$ time.

The coordination gain of the periodic drive is a separate finding. At
equal total interaction strength the grid and the ring have the same
transverse tilt yet the grid lives $2.15$ times longer at $N=12$ and
$2.78$ times at $N=16$, so coordination acts through something the tilt scaling does not contain. A gain from added neighbours is expected in
direction, since a one-dimensional short-range system lacks the
stabilisation that higher dimensions can supply~\cite{else2017prethermal}.
Our results say nothing about its size in the thermodynamic limit. The
gain depends on $\gamma$, grows between $N=12$ and $N=16$, and is not
explained by our picture. For the multipolar drives we established only
directions at fixed $zJ$, an excess for $n=2$ and a definition-dependent
sign for Thue-Morse.

The limits are as follows. The systems are small ($N=12$ and $16$, on single lattices, with the $N=12$ grid containing triangles that the $N=16$ grid does not), and we do not extrapolate. The disorder strength, the pulse
error and the form of dephasing are fixed, apart from the scans reported. The initial state is fixed (fully polarised).
The toggling-frame estimates are first order in the pulse error and
uncontrolled at the operating point, and the closed form of
Eq.~\eqref{eq:block_resid} fails for the absolute $n=2$ time. The coherence
time is a threshold time whose value depends on the definition, which we
have tracked by reporting three, and it is not an
energy-absorption rate. The ring at $J=0.1$ has no locked plateau, so
ratios against it are quoted only in tables. The Thue-Morse response
depends strongly on $\gamma$ and we do not explain it.

For larger systems we tested the ordering of the random multipolar orders on
a one-dimensional open chain of $N=50$ sites by time-evolving block
decimation with quantum-jump dephasing~\cite{daley2014,vidal2004}. The parameters were $J=0.1$, $D=1.5$, $g=0.9$, per-site dephasing $\gamma=0.01$,
bond dimension $\chi=64$, $8$ disorder seeds and $2$ trajectories per seed,
with an initial window of $300$ steps that was doubled until no seed was pinned. Table~\ref{tab:mps} gives $T_c$ (last exceedance) and, for
comparison, the ED ring at $N=12$. The largest truncation error at $T_c$ over
seeds is $0.119$, $0.167$ and $0.242$ for $n=0$, $1$ and $2$. Doubling the
bond dimension to $\chi=128$ for $n=2$ changes the mean from $108.75$ to
$108.50$. The ordering $n=0<n=1<n=2$ of the $N=12$ ring is reproduced at
$N=50$. The periodic drive is not reported at $N=50$ because its bond
dimension is not converged, and Thue-Morse was not run. The calculation is
one-dimensional and does not address coordination.

\begin{table}[!htb]
\caption{Matrix-product-state coherence times for the random multipolar drives on an open chain of $N=50$ sites ($J=0.1$, $D=1.5$, $g=0.9$, per-site dephasing $\gamma=0.01$, bond dimension $\chi=64$, $8$ seeds, $2$ quantum-jump trajectories per seed, initial window $300$ steps), and the ED ring at $N=12$ ($J=0.1$, periodic, $48$ seeds).}
\label{tab:mps}
\begin{ruledtabular}
\begin{tabular}{lcc}
Drive & MPS, $N=50$ & ED ring, $N=12$ \\
\hline
$n{=}0$ & $25.25\,(3.66)$ & $26.5\,(1.5)$ \\
$n{=}1$ & $74.00\,(9.41)$ & $72.6\,(4.0)$ \\
$n{=}2$ & $108.75\,(8.82)$ & $99.0\,(6.1)$ \\
\end{tabular}
\end{ruledtabular}
\end{table}

\FloatBarrier
\section{Conclusion}
\label{sec:conclusion}

We have studied, by exact diagonalisation, how interactions affect
subharmonic order in a disordered kicked Ising model under periodic
and multipolar drives, and separated the role of coupling strength
from that of coordination. The response to interaction strength
reverses with the drive. Interactions protect a periodically driven time
crystal against dephasing by reducing the transverse tilt on which dephasing
acts, and destabilise multipolar-driven order of order $n\ge1$ by spoiling
the cancellation of the pulse error. At equal total interaction strength a
grid outlasts a ring for the periodic drive by $2.15$ ($N=12$) and $2.78$
($N=16$), although the transverse tilt is the same, a coordination effect
beyond the toggling-frame picture, and for $n=2$ in direction. Doubling the
coupling on a ring shortens the $n=1$, $n=2$ and Thue-Morse times to
$0.54$, $0.37$ and $0.44$ at $N=16$. These results rest on exact diagonalisation of small systems and
on three definitions of the coherence time, and are organised, but not
fully explained, by a toggling-frame picture.

\begin{acknowledgments}
The authors thank the Photonics and Quantum Technology Laboratory,
PES University, for computational resources and support.
\end{acknowledgments}

\section*{Data Availability}
The data and code supporting the findings of this study are available
from the corresponding author upon reasonable request.

\bibliographystyle{apsrev4-2}
\bibliography{references}

\end{document}